\documentclass[draftclsnofoot, onecolumn, 12pt]{IEEEtran}
\usepackage{upgreek}
\usepackage[caption=false,font=normalsize,labelfont=sf,textfont=sf]{subfig}
\usepackage{array}
\usepackage{stfloats}
\usepackage{url}
\usepackage{verbatim}
\usepackage{colortbl}
\usepackage{diagbox}
\usepackage{adjustbox}
\usepackage{float}
\usepackage{lipsum}
\usepackage{boldline}
\usepackage{hhline}
\usepackage[dvipsnames]{xcolor}
\usepackage[group-separator={,},group-minimum-digits=4]{siunitx}

\usepackage{graphicx,epsfig}
\usepackage[noadjust]{cite}
\usepackage{mcite}
\usepackage{amsfonts,helvet}
\usepackage{fancyhdr}
\usepackage{threeparttable}
\usepackage{epsf,epsfig}
\usepackage{amsthm}
\usepackage{amsmath}
\usepackage{amssymb}
\usepackage{dsfont}
\usepackage{color}
\usepackage[linesnumbered,ruled]{algorithm2e}
\usepackage{enumerate}
\usepackage{cancel}
\usepackage{comment}
\usepackage{booktabs}
\usepackage{multirow}
\usepackage{gensymb}

\SetKwInput{KwInput}{Input}
\SetKwInput{KwInt}{Initialize}
\SetKwInput{KwOutput}{Output}
\SetKwProg{Fn}{Stage 1}{}{}
\SetKwProg{Fn}{Stage 2}{}{}
\SetKw{Continue}{continue}

\usepackage{eucal}
\usepackage{stix,bbding,pifont,utfsym,fontawesome}

\def\xmark{\textcolor{red}{\XSolidBrush}}
\def\tickmark{\textcolor{green}{\ding{52}}}

\usepackage{rotating}
\usepackage{makecell}

\begin{document}
\title{{Beyond GNSS: Leveraging LEO Satellite Signals of Opportunity for Doppler  Positioning }}
\author{Md. Ali Hasan, M. Humayun Kabir, 
Sangmin Han, Wonjae Shin,  H. Vincent Poor

\thanks{Md. Ali Hasan, Sangmin Han, and Wonjae Shin are with the School of Electrical Engineering, Korea University, Seoul 02841, Republic of Korea (e-mail: \texttt{alihasan@korea.ac.kr; smhan22@korea.ac.kr; wjshin@korea.ac.kr}).}
\thanks{M. Humayun Kabir is with the Department of Electrical and Electronic Engineering, Islamic University, Kushtia 7003, Bangladesh (e-mail: \texttt{humayun@eee.iu.ac.bd}).}
\thanks{H. V. Poor is with the Department of Electrical and Computer Engineering, Princeton University, Princeton, NJ 08544, USA (e-mail: {\texttt{poor@princeton.edu}}).}
}
%
\maketitle 

\begin{abstract}

\label{sec:abstract}Large constellations of low Earth orbit (LEO) satellites provide ubiquitous global connectivity; yet, their rapid motion introduces significant Doppler shift effects. 
While challenging for communication, these effects can be leveraged for precise user terminal (UT) positioning, particularly in GNSS-denied environments where traditional satellite-based navigation struggles. Unlike GNSS, most LEO satellites lack dedicated navigation signals, motivating signals-of-opportunity (SOPs) approaches that repurpose existing LEO signals and use publicly available two-line element (TLE) files for orbital information, although ephemeris errors can limit accuracy. 
 This article explores the fundamental principles, challenges, and enabling technologies for LEO-based Doppler shift positioning, highlighting its advantages over conventional non-Doppler methods, and discussing future research directions. Furthermore, to facilitate realistic and reproducible performance evaluation beyond stochastic modeling, we propose \mbox{SionnaNTN-RT}, a ray tracing framework for non-terrestrial network (NTN) built upon Sionna, enabling end-to-end simulation of Doppler-based positioning. Simulation results demonstrate that Doppler-based positioning can achieve high positioning accuracy in both urban and open-sky environments, highlighting its promise for next-generation positioning systems.

\end{abstract}

\begin{IEEEkeywords}
Doppler shift-based positioning, Differential Doppler positioning, Ephemeris error, LEO satellite, Signals of opportunity (SOPs). 
\end{IEEEkeywords}

\section{Introduction}
\label{sec: Introduction}

{The ability to accurately determine position is essential for delivering user-centric services across various domains, including mobile communications, vehicular networks, precision agriculture, direct‑to‑device (D2D) communication, and defense applications.}
Over the past few decades, the Global Navigation Satellite System (GNSS) has been the dominant technology for outdoor positioning that utilizes satellites in medium Earth orbit (MEO). Despite its widespread adoption, GNSS faces significant limitations in certain environments.
Due to the high altitude, GNSS signals' strength is insufficient in urban canyons and other challenging conditions, such as a limited number of visible satellites, making pseudorange-based positioning unreliable.
To address GNSS outages, several assisted positioning solutions—such as distance measuring equipment (DME), pseudolite networks, and passive wide-area multilateration (P-WAM) \cite{7593455}—have been proposed.
These solutions, while effective in specific regions, require dedicated infrastructure and hardware, limiting their scalability.
Doppler shift-based positioning has emerged as a viable alternative, as Doppler measurements remain reliable even in harsh environments without requiring carrier signal demodulation.
However, its application in GNSS has been limited due to poor Doppler dilution of precision (DDOP), a key metric quantifying the impact of satellite velocity direction and geometric distribution on positioning accuracy. 
The low angular motion and high altitude of the GNSS satellites result in poor DDOP values, leading to degraded positioning accuracy \cite{shi2023revisiting}. 
In contrast, low Earth orbit (LEO) satellites, like Starlink, have higher angular motion and lower orbital altitude, producing more substantial Doppler shift effects, which enable more accurate positioning.

The rapid proliferation of large-scale LEO satellite constellations has sparked interest in developing complementary or stand-alone Positioning, Navigation, and Timing (PNT) solutions beyond traditional GNSS.
While primarily designed for communication, these LEO satellites emerge as a promising platform for high-accuracy positioning applications \cite{10388052}.
R. M. Ferre \textit{et al.} \cite{9800136} explore the use of LEO satellites for future equal-access localization; however, Doppler-based or differential positioning is not considered.
Unlike GNSS, most LEO satellites do not broadcast the navigation message, and adopting them for positioning would require protocol modifications, whose feasibility depends on satellite operators' intentions \cite{10355106}.
To leverage existing LEO constellations for PNT, researchers have explored a signals-of-opportunity (SOPs) approach, which utilizes
radio signals from LEO satellites
and extracts ephemeris data from two-line element (TLE) files. 
The CelesTrak website updates the TLE files every 8-hour cycle, and the UT can get it upon availability of a communication gateway.
This approach provides viable positioning alternatives, albeit with accumulated ephemeris errors, often mitigated through inertial navigation system (INS)-based ephemeris tracking \cite{10388052}, which increases hardware complexity.
Table~\ref{tab: paper_com} compares existing magazine articles in terms of their coverage of key aspects of Doppler-based positioning.
Most existing studies focus on hybrid positioning, generally overlooking DDOP analysis, and providing limited or no evaluation of positioning performance. In contrast, our article presents GNSS- and INS-independent Doppler-based methods, offering detailed DDOP analysis and comprehensive positioning performance evaluation.

\begin{table}[b]\renewcommand{\arraystretch}{1}
\centering
  \caption{Comparison of existing magazine articles related to LEO Doppler-based positioning}
\resizebox{0.9\linewidth}{!}{
    \begin{tabular}
    {>{\hspace{-1.8em}}c<{\hspace{-.2em}} 
    >{\hspace{-1.85em}}c<{\hspace{-.7em}}
    >{\hspace{-1.51em}}c<{\hspace{-1.3em}}
    >{\hspace{-1.1em}}c<{\hspace{-1.2em}}
    >{\hspace{-1.1em}}c<{\hspace{-.7em}}
    >{\hspace{-1.5em}}c<{\hspace{-.7em}}
    >{\hspace{-1.5em}}c<{\hspace{-.7em}}
    >{\hspace{-1.23em}}c<{\hspace{-.2em}}
    >{\hspace{-1.5em}}c<{\hspace{-1.56em}}}
    \hlineB{3}
         \begin{tabular}{c} Positioning \\ Signal \end{tabular}  &Ref. &  \begin{tabular}{c} Doppler \\ Positioning \end{tabular} & \begin{tabular}{c} LEO \\ Satellite  \end{tabular}& \begin{tabular}{c} Differential \\ Positioning \end{tabular} 
         & \begin{tabular}{c} GNSS \\ Free \end{tabular}
         & \begin{tabular}{c} INS \\ Free \end{tabular}
         &DDOP & \begin{tabular}{c} Performance \\ Evaluation \end{tabular}  \\
         \hlineB{3}
          & ~\cite{7593455} & \xmark &\xmark &\xmark &\xmark &\tickmark &\xmark &\xmark  \\ 
          \multirow{-2}{*}{ \begin{tabular}{c} \cellcolor[HTML]{FFFFFF} ~~~PNT~~ \\ \cellcolor[HTML]{FFFFFF} ~Signal \end{tabular} } & \cellcolor[HTML]{EEF3FF} \cite{10355106} &\cellcolor[HTML]{EEF3FF}\xmark &\cellcolor[HTML]{EEF3FF}\tickmark &\cellcolor[HTML]{EEF3FF}\xmark &\cellcolor[HTML]{EEF3FF}\tickmark &\cellcolor[HTML]{EEF3FF}\tickmark &\cellcolor[HTML]{EEF3FF}\xmark &\cellcolor[HTML]{EEF3FF}\tickmark\\
         \hline
         & ~\cite{9800136}  &\xmark &\tickmark &\xmark &\tickmark &\tickmark &\xmark &\xmark \\ 
         &\cellcolor[HTML]{EEF3FF} \cite{10388052}  &\cellcolor[HTML]{EEF3FF}\tickmark &\cellcolor[HTML]{EEF3FF}\tickmark &\cellcolor[HTML]{EEF3FF}\xmark &\cellcolor[HTML]{EEF3FF}\xmark &\cellcolor[HTML]{EEF3FF}\xmark &\cellcolor[HTML]{EEF3FF}\xmark &\cellcolor[HTML]{EEF3FF}\tickmark \\
         \multirow{-3}{*}{\begin{tabular}{c} ~~~SOPs \\ ~~~Approach \end{tabular} } &This work  &\tickmark &\tickmark &\tickmark &\tickmark &\tickmark &\tickmark &\tickmark \\   
         \hlineB{3} \noalign{\vskip 0.2em}
          \multicolumn{8}{l}{
        \begin{tabular}{l l l r}
             \tickmark ~~topic addressed &&&
             \xmark ~~topic not addressed  
        \end{tabular}}
    \end{tabular}
    }
    \label{tab: paper_com}
\end{table}

Doppler shift-based positioning offers several advantages over traditional pseudorange-based positioning when using LEO satellites.
Code-phase pseudorange estimation remains challenging due to asynchronous clocks among LEO satellites and the absence of dedicated PNT signals.
On the other hand, carrier phase-based pseudorange estimation, while highly accurate, presents additional challenges due to the complexity of carrier ambiguity resolution.
In contrast, Doppler shift-based positioning eliminates these drawbacks, making it a more practical alternative for real-world deployment.
%
%
Despite its potential, Doppler shift-based positioning using LEO satellites faces several technical challenges that must be addressed.
Unlike GNSS, LEO satellites lack precise clock synchronization, leading to errors in Doppler shift measurements. 
{Additionally, ephemeris uncertainty is another source of error; the rapid movement of LEO satellites requires continuous updates for reliable position estimation.}
One potential solution is differential positioning, where a base receiver provides corrections to mitigate these errors \cite{dif9843493}, while dedicated ephemeris correction algorithms can further enhance positioning accuracy \cite{10716274}.
{By mitigating these challenges, Doppler positioning can serve as a robust alternative to existing satellite-based methods in environments where GNSS performance is compromised.}

The main contributions of this article are as follows:
\begin{itemize}
    \item 
Motivated by the growing potential of Doppler shift-based positioning enabled by large-scale LEO constellations, this article provides a comprehensive tutorial and review of its fundamental principles, key challenges, and potential applications.  
    \item
We propose \textit{\mbox{SionnaNTN-RT}}, a non-terrestrial network (NTN) ray-tracing framework built upon Sionna that integrates a beamforming-based mechanism for efficient ray candidate generation. 
By incorporating real satellite ephemerides, geographic environments, and signal propagation conditions, \mbox{SionnaNTN-RT} enables reproducible and realistic evaluation of Doppler-based positioning.
%
%
%
    \item 
Using \mbox{SionnaNTN-RT}, we comparatively evaluate the positioning performance of representative Doppler and non-Doppler methods in open-sky and urban environments and discuss future research directions to enhance the accuracy and reliability of LEO-based positioning systems.
\end{itemize}

        

\section{Key Benefits of LEO Satellites for Positioning}
 
Most existing LEO satellites are not designed for navigation functionality and do not broadcast positioning signals; enabling navigation support may therefore incur service-related costs.
In contrast, emerging satellite providers, such as Xona and CentiSpace, are developing dedicated navigation signals within their LEO constellations, offering potential alternatives to traditional GNSS.
Fig.~\ref{fig: DDOP} presents a summary of LEO satellite constellations, along with GNSS.
The potential benefits of the LEO constellation are described below.


\subsubsection{Large Constellation}

Recent advances in satellite launch technology have facilitated large-scale LEO constellation deployment, providing broadband connectivity for unconnected regions on Earth.
Currently, Starlink operates $\num{10725}$ satellites, with plans to expand to $\num{41943}$ shortly.
This large-scale deployment ensures continuous global coverage, significantly alleviating positioning disruptions due to satellite outages.

\subsubsection{High SNR}

Compared with GNSS satellites, LEO satellites orbit approximately twenty times closer to Earth, resulting in significantly stronger received signals. The high signal-to-noise ratio (SNR) of the received signal enhances the resilience of the UT against jamming, spoofing, and multipath effects. 
Additionally, the stronger signal allows UTs to receive LEO satellite signals in indoor and highly obstructed environments \cite{9800136}, reducing errors in Doppler shift measurements.

\subsubsection{High Satellite Visibility}

Large LEO constellations ensure a greater number of visible satellites; for instance, with full Starlink deployment, up to $\num{344}$ satellites may be visible to a UT simultaneously \cite{9800136}.
The high satellite density mitigates the risk of signal outages, improving positioning reliability compared to GNSS, which often suffers from satellite blockages in urban environments. 

\subsubsection{Improved Geometric Diversity}

LEO satellites orbit at altitudes of approximately $\num{600}$ km, with $\num{90}$-minute orbital periods and $7.56$ km/s velocity.
This rapid movement continuously changes satellite-to-user geometry, improving positioning accuracy.
Unlike GNSS, where blockages between the UT and satellites can persist for long durations, LEO satellites move quickly, reducing the impact of line-of-sight (LOS) obstructions.

\begin{figure*}[t]
    \centering
    \includegraphics[width=\linewidth]{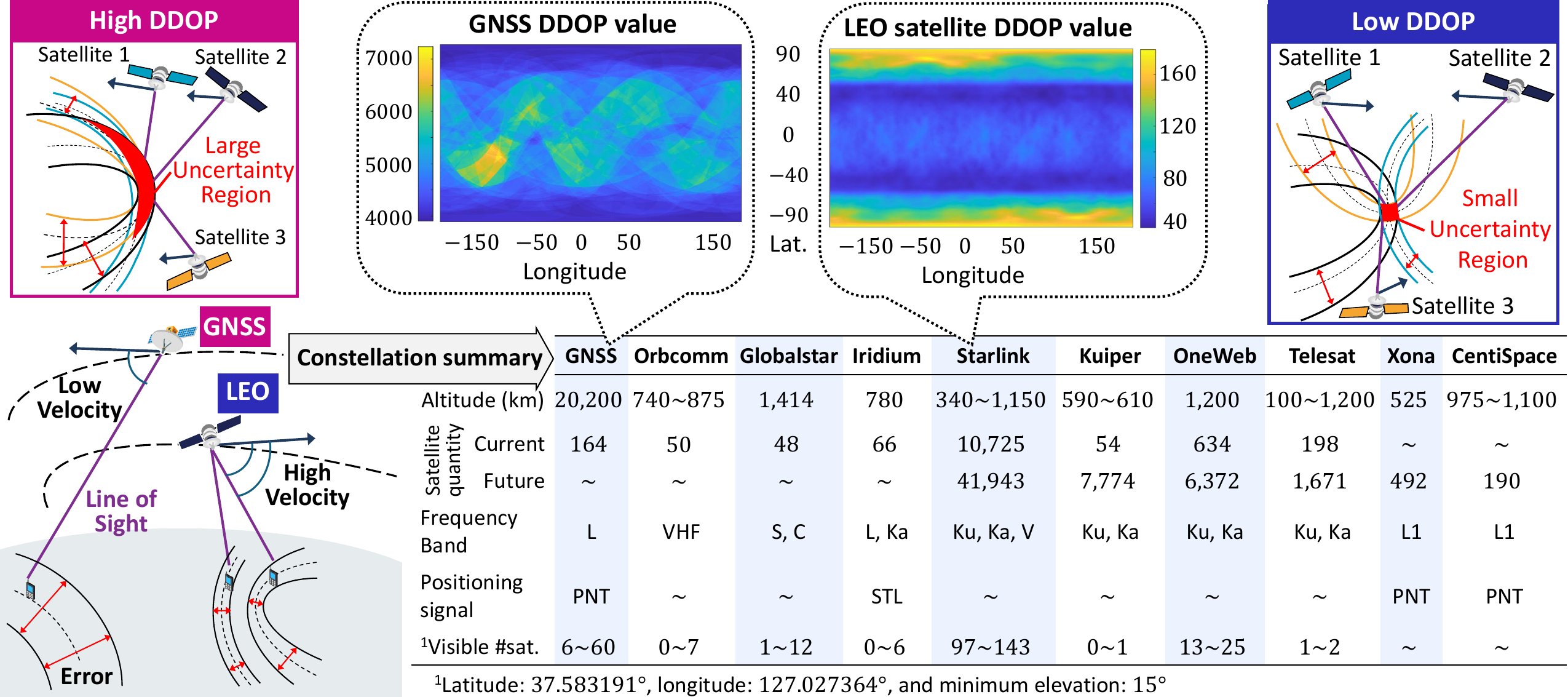}
    \caption{DDOP illustration and heatmap for Starlink and GNSS with summary of satellite constellations.}
    \label{fig: DDOP}
\end{figure*}


\section{Doppler Shift-Based Positioning with LEO}

While Doppler-based positioning offers limited accuracy in GNSS, the pronounced Doppler shifts of LEO satellites make it a promising solution for UT positioning.
This section discusses the motivations, fundamental principles, and challenges of Doppler shift-based positioning using LEO satellites.

\subsection{Motivations}
GNSS signals are highly susceptible to environmental impairments, leading to ephemeris acquisition failures and degraded UT positioning accuracy \cite{Spensnavi.537}.
In contrast, LEO satellite signals provide improved signal availability, making Doppler-based positioning a promising alternative.
Moreover, the Doppler measurements from the LEO satellite offer several unique advantages over pseudorange-based positioning.

 
\subsubsection{High Positioning Accuracy}


Measurement errors significantly impact positioning performance; Doppler shift can be precisely estimated directly from the received signal without demodulating the carrier signal \cite{gowdayyanadoddi2015ray}, enabling LEO-based Doppler positioning to outperform commercial-grade GNSS receivers, particularly in challenging environments such as indoor and urban canyons.

\subsubsection{SOPs-Based Approach}

The SOPs approach utilizes existing radio signals that were not originally designed for navigation. 
Among the three major satellite-based navigation observables—pseudorange, carrier phase, and Doppler shift—Doppler shift is the most practical for SOPs-based positioning. The UT can extract the Doppler shift by simply measuring the frequency difference between the transmitted and received signals.
Carrier phase tracking, on the other hand, suffers from cycle slips, which introduce additional errors.

\subsubsection{Less Environmental Effect}

Positioning observables, which capture key measurements for localization, are significantly affected by environmental conditions and signal propagation characteristics. 
In urban and indoor environments, satellite signals are blocked or received under non-LOS (NLOS) conditions.
Pseudorange-based positioning, which depends on signal propagation time, is highly susceptible to multipath interference, making it unreliable in such conditions.
Doppler shift-based positioning, however, relies on frequency changes due to the relative motion between satellite and UT, making it much less affected by multipath errors \cite{gowdayyanadoddi2015ray}.

\subsubsection{No Ambiguity Resolution}

The Doppler shift is the time derivative of the carrier phase.  
{However, in carrier phase-based positioning, an unknown number of full-wavelength cycles, known as carrier ambiguity, must be resolved to achieve accurate positioning, which is computationally intensive and introduces additional errors \cite{9110199}.}
In contrast, Doppler shift-based positioning eliminates the need for ambiguity resolution, allowing for instantaneous position estimation.

\subsubsection{Low DDOP}

The accuracy of Doppler shift-based positioning is closely related to DDOP, a metric that quantifies the impact of satellite velocity direction and geometric diversity on positioning accuracy.
DDOP depends on satellites' altitude, velocity, acceleration, and satellite-to-user geometry. 
Lower DDOP values indicate better positioning accuracy. 
GNSS satellites, with their high altitude and relatively low velocity, have DDOP of order $10^4$, resulting in poor positioning precision. In contrast, LEO satellites exhibit a DDOP on the order of $10^2$, ensuring more reliable positioning opportunities \cite{shi2023revisiting}.
Fig.~\ref{fig: DDOP} illustrates the DDOP value of visible Starlink LEO satellites compared to those of GNSS satellites, confirming that LEO-based positioning offers consistently lower DDOP values across large portions of the Earth.


\subsection{Fundamental Positioning Principles}

\begin{figure*}
    \centering
    \includegraphics[width=\linewidth]{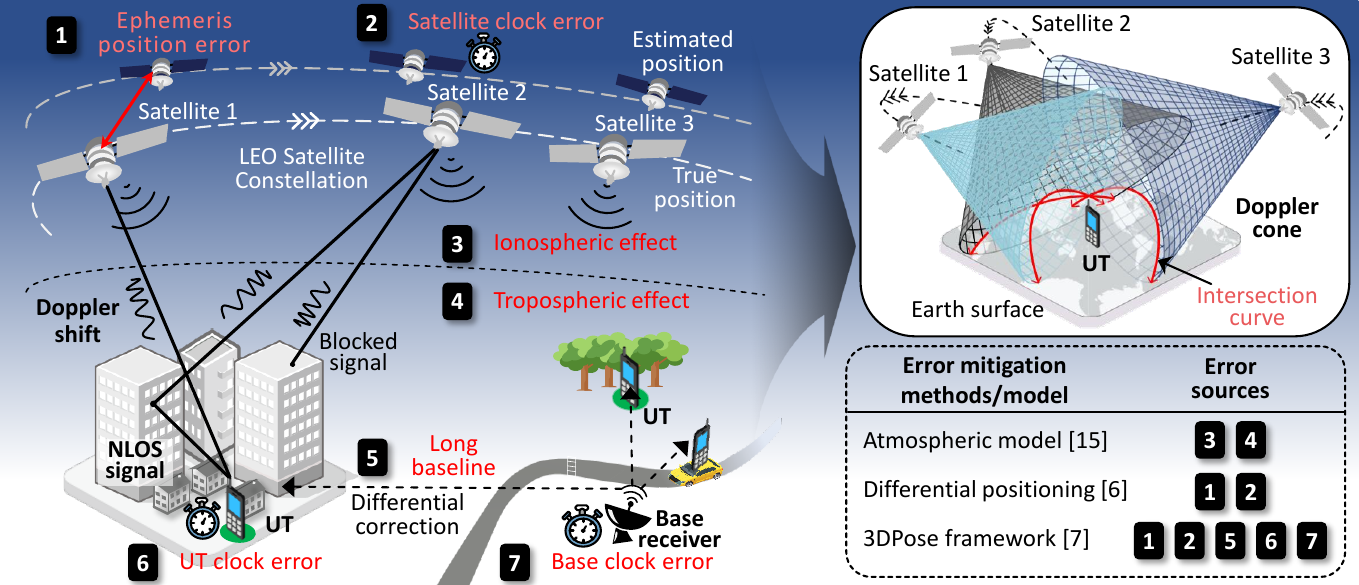}
    \caption{Doppler shift-based positioning with LEO satellites: positioning geometry, error sources, and mitigation methods.}
    \label{fig: scenario}
\end{figure*}

{The Doppler shift refers to frequency changes of the transmitted signal due to the relative velocity between the transmitter (satellite) and the receiver (UT) along the LOS direction, forming the basis for Doppler shift-based positioning.}
Specifically, the UT estimates its position by calculating the pseudorange rate, which is the time derivative of the pseudorange between the satellite and UT, from Doppler shift measurements, where the pseudorange rate is computed as the negative product of the Doppler shift and the transmitted signal wavelength. 
The higher the radio carrier frequency, the better the positioning accuracy, as it produces larger Doppler measurements for a given relative velocity.
A receiver architecture for reliably acquiring and tracking Doppler shifts from unknown LEO satellites is detailed in \cite{11165366}.
For a stationary UT, the satellite-to-user LOS vector forms an angle with the satellite velocity direction, creating a right circular cone with the satellite position as the vertex. This geometric structure is known as the {\it Doppler cone}, and its surface represents the set of possible UT positions. 
The intersection of multiple Doppler cones with Earth's surface, as illustrated in Fig.~\ref{fig: scenario}, determines the precise UT position. 
The axis vector of the right circular cone is collinear with the satellite velocity vector, acting in the same and opposite direction for a positive and negative Doppler shift, respectively.
By analyzing the intersection of multiple Doppler cones, the UT's correct position can be estimated with high accuracy.

\subsection{Key Challenges}


\subsubsection{Clock Synchronization}
 
Unlike GNSS, most LEO satellites lack high-precision atomic clocks, resulting in asynchronous clock drift across satellites.
This drift introduces transmit frequency deviation that causes frequency instability, contaminating Doppler shift measurements, which must be accounted for in the positioning algorithm.
Since each satellite has an independent clock drift, the number of unknown clock errors scales with the number of measurements, making accurate clock estimation essential to improve positioning accuracy.

\subsubsection{Ephemeris Error}

LEO satellites do not necessarily broadcast orbital parameters, as they are primarily designed for broadband communication rather than navigation.
Instead, ephemeris data is commonly derived from infrequently updated TLE files, leading to accumulated ephemeris errors over time.
Studies such as \cite{10288195} have reported satellite position and velocity errors of approximately $\num{3}$ km and $\num{3}$ m/s, respectively.
These ephemeris errors translate into significant UT positioning errors, often reaching several kilometers \cite{shi2023revisiting}, which need to be addressed to improve positioning accuracy.

\subsubsection{Receiver Design}

The receiver design for the UT plays a critical role in PNT solutions, responsible for signal acquisition, tracking, and extracting key measurements such as Doppler shift. 
The accuracy of the positioning solution is largely dependent on the receiver's ability to process these measurements effectively.
While commercial-off-the-shelf GNSS receivers are widely available, dedicated PNT receivers utilizing LEO constellations are not yet commercially accessible.
Additionally, most LEO satellite signals are proprietary, as satellite operators do not publicly disclose signal structures.  
This lack of transparency complicates the development of LEO-PNT receivers, constraining hardware and software design from effectively utilizing these signals.

\subsubsection{Satellite Identification}
  
Unlike GNSS, which broadcasts dedicated navigation messages, LEO satellites do not inherently provide identification information in their transmitted signals.
Consequently, UTs must infer satellite identities using Doppler shift measurements and prior orbital knowledge, which becomes difficult with many visible satellites, among thousands of active satellites.
In light of this, developing a robust satellite identification method is essential for improving positioning accuracy. 
Possible approaches include using auxiliary network information to narrow down visible satellites or applying machine learning (ML) algorithms for predicting satellite identity from observed signal characteristics.

\subsubsection{Algorithmic Convergence}

The relationship between Doppler shift and UT position is inherently nonlinear, which requires linearizing the Doppler shift equation around an initial position guess and progressively refining the estimate through iterative computation \cite{10716274}.
This process forms a four-column gradient matrix, representing the three UT position coordinates and one for clock drift.
The positioning algorithm may diverge if initial position error exceeds a certain threshold, such as $\num{410}$ km \cite{shi2023revisiting}.
Expanding the gradient matrix to seven columns by including velocity as an additional unknown improves convergence reliability regardless of initial error, but requires enough visible satellites to ensure a well-conditioned solution.

\subsection{Key Enabler Technologies}
Despite its challenges, Doppler shift-based positioning with LEO satellites holds great potential for both stand-alone and augmented positioning systems. Several enabling technologies can enhance its feasibility and accuracy.


\subsubsection{Stand-alone Positioning}

While LEO satellites are primarily designed for broadband internet service, their signals can be repurposed for positioning through the SOPs approach. 
By extracting Doppler shift information, a stand-alone positioning solution can be achieved without requiring dedicated PNT signals.
The technique involves superimposing multiple three-dimensional (3D) Doppler cones to determine the UT position.
However, the accuracy of this method relies heavily on the availability of precise satellite ephemeris data.

\subsubsection{GNSS Augmentation with LEO Satellites}

GNSS-based positioning suffers in challenging environments such as urban canyons and dense foliage, where limited satellite visibility degrades accuracy. 
The strong signals from LEO satellites provide an opportunity to augment GNSS. By integrating LEO Doppler shift measurements with GNSS data, orbital errors in LEO satellites can be estimated and corrected. This hybrid approach ensures improved positioning reliability, even during GNSS outages.

\subsubsection{Data-Driven Approach}

ML and data-driven approaches hold promise for enhancing LEO-based positioning. 
Beamforming-based fingerprinting, where received signal strength indicators (RSSI) patterns are collected from different locations, can be used to train ML models for localization. 
Furthermore, automated ML techniques can refine orbit prediction models, improving the accuracy of satellite ephemeris.

\subsubsection{INS Fusion with Doppler Positioning}

An INS is a self-sufficient navigation system that can provide navigation information in the local coordinate frame, though it accumulates error over time.
On the other hand, Doppler shift-based positioning with LEO satellites suffers from ephemeris error, which causes significant UT position errors. The integration of INS and Doppler shift can leverage the strengths of both systems, improving overall positioning accuracy.

\subsubsection{Assisted Doppler Positioning}

Similar to Assisted GNSS (A-GNSS), Assisted Doppler positioning (A-DP) can improve accuracy by providing real-time atmospheric and ephemeris corrections to the UT via an Internet connection. 
Integrated base stations within TN can deliver updated TLE files and error corrections to the UT, significantly enhancing positioning performance.

\subsubsection{Differential Doppler Positioning}

To mitigate satellite-related errors, the differential positioning technique introduces a base receiver with a precisely known location. This base station measures Doppler shifts and estimates correction factors, which are then applied to UT measurements to reduce positioning errors. 
This method's effectiveness largely depends on the spatial proximity between the UT and the base station, as errors tend to increase with longer baselines.

\subsubsection{Positioning with Ephemeris Error Correction}

Ephemeris errors accumulate over time due to the reliance on outdated TLE data.
The proposed {\it 3DPose} framework \cite{10716274} addresses this issue by modeling and compensating for the effects of ephemeris errors through a novel ephemeris error correction algorithm. 
This algorithm explicitly characterizes and corrects Doppler-shift measurements by accounting for satellite position errors in the tangential direction, thereby significantly improving UT positioning accuracy.


\section{Results and Discussion}

This section presents a comparative analysis of Doppler and non-Doppler positioning methods in open-sky and urban environments to evaluate their positioning accuracy and robustness under different propagation conditions.
The evaluation focuses on 3D positioning root mean square error (RMSE), time series error, and evaluation maps across four distinct scenarios, combining static and moving UTs in each environment. 
In an open-sky environment, there is no multipath effect, whereas the urban environment introduces significant multipath errors and NLOS effects due to reflections and obstructions from high-rise buildings, potentially biasing the pseudorange measurement by up to $\num{100}$ m.
Our simulation framework incorporates 3D building structures from OpenStreetMap, while the multipath errors are estimated using the proposed \mbox{SionnaNTN-RT} ray-tracing for realistic evaluation.

\subsection{SionnaNTN-RT Ray Tracing}

Sionna is an open-source TensorFlow library for wireless communication research that supports differential ray-tracing \cite{hoydis2022sionna}.
While it has been widely used for terrestrial networks (TN), its extension to NTN remains largely unexplored.
Sionna supports 3D environment modeling through the Mitsuba-3 rendering framework and the just-in-time compiler Dr.Jit.
Ray-tracing involves shooting and bouncing rays (SBR) methods that initialize candidate rays in all directions of a unit sphere. However, this approach is unsuitable for the NTN scenarios, as the satellite signals are primarily directed toward the Earth's surface. To address this limitation, we propose \mbox{{\it SionnaNTN-RT}}, a ray-tracing framework tailored for NTN, which modifies the candidate ray generator to employ a beamforming strategy that directs all (approximately one million) candidate rays toward the target environment. The \mbox{SionnaNTN-RT} can generate the LOS and NLOS signal propagation channels with the satellite's real position. 
Instead of stochastic modeling of the wireless channel, \mbox{SionnaNTN-RT} generates deterministic channel parameters relying on physical modeling of the real-world environment.
The multipath effect on the Doppler shift is computed utilizing a power-weighted sum of Doppler shifts from all signal propagation paths.

\begin{figure*}[t]
    \centering
    \includegraphics[width=\linewidth]{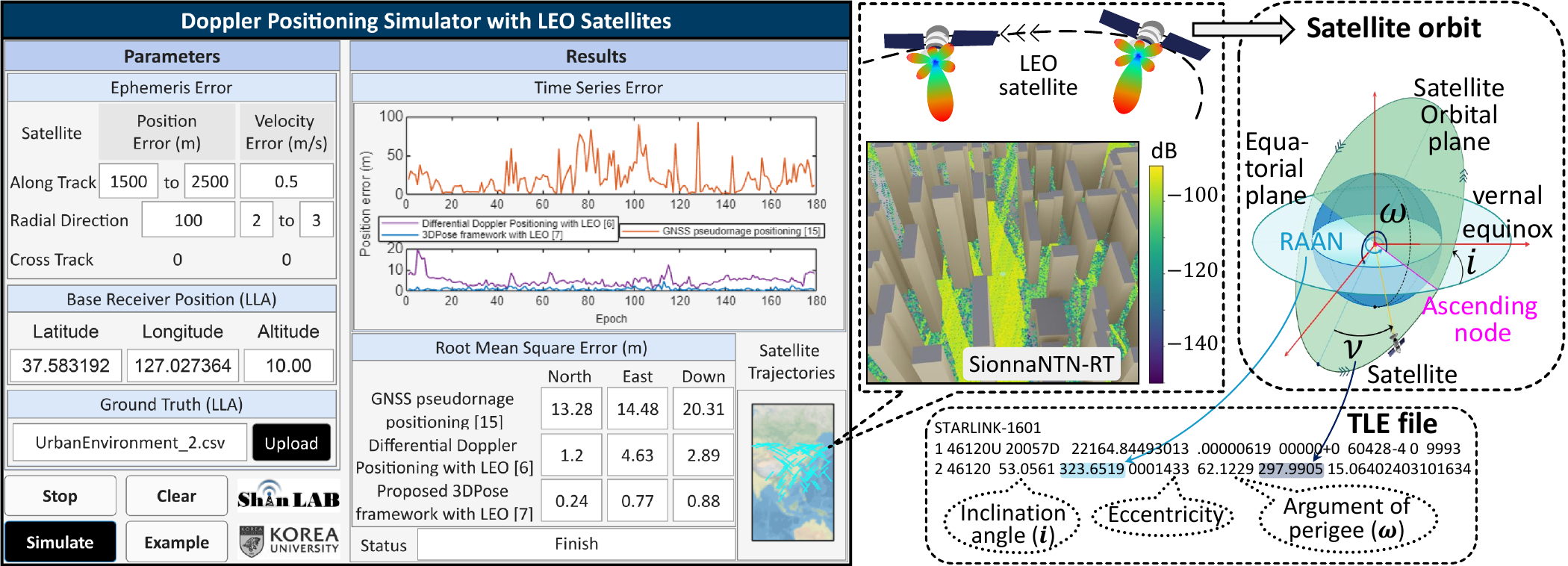}
    \caption{LEO satellite-based Doppler positioning GUI simulator and satellite orbit. By varying the satellite position and velocity errors, the simulator evaluates the impact of ephemeris errors on Doppler positioning and visualizes the resulting positioning performance for differential Doppler, the 3DPose framework with LEO satellites, and GNSS pseudorange positioning.}
    \label{fig: simulator}
\end{figure*}

\subsection{Positioning Error Analysis}
The positioning methods are categorized into Doppler and non-Doppler approaches, which are further classified as non-differential and differential methods.
Non-differential methods rely solely on UT measurements, whereas differential positioning introduces a base receiver with a known position to improve accuracy.
To evaluate the Doppler shift-based positioning, it is compared with carrier phase-based and GNSS pseudorange-based positioning. The non-differential positioning methods considered in this evaluation include non-differential Doppler positioning \cite{shi2023revisiting}, Doppler positioning based on extended Kalman filter (EKF) \cite{BENZERROUK2019496}, and GNSS pseudorange positioning \cite{morton2021position}. The differential positioning methods include double-difference carrier phase positioning \cite{9110199}, differential Doppler positioning \cite{dif9843493}, and 3DPose framework \cite{10716274}.  
For a consistent comparison, we implement and evaluate these positioning frameworks using the same simulated measurement dataset.
The Doppler shift and carrier phase calculations use MATLAB's Satellite Communications Toolbox with TLE files for $118$ Starlink satellites, while GNSS pseudorange calculations use TLE files of $60$ GNSS satellites.

 
Specifically, 
the Starlink ephemerides derived from TLEs using the SGP4 orbit propagator model exhibit errors of approximately $\num{3}$ km in position and $\num{3}$ m/s in velocity.
Position errors are mainly concentrated in the tangential direction, ranging from $\num{1.5}$ to $\num{2.5}$ km, while radial errors are around $\num{200}$ m. Likewise, velocity errors range from {$\num{2}$ to $\num{3}$ m/s in the radial direction and $\num{0.5}$ m/s in the tangential direction. 
These error characteristics are randomly incorporated into our simulation, assuming initial position and velocity errors drawn from a uniform distribution.
In addition to the satellite-based errors, tropospheric and ionospheric errors are included in the Doppler shift calculation using the Saastamoinen and Klobuchar models, respectively \cite{morton2021position}.

\begin{figure*}[t]
    \centering
    \includegraphics[width=\linewidth]{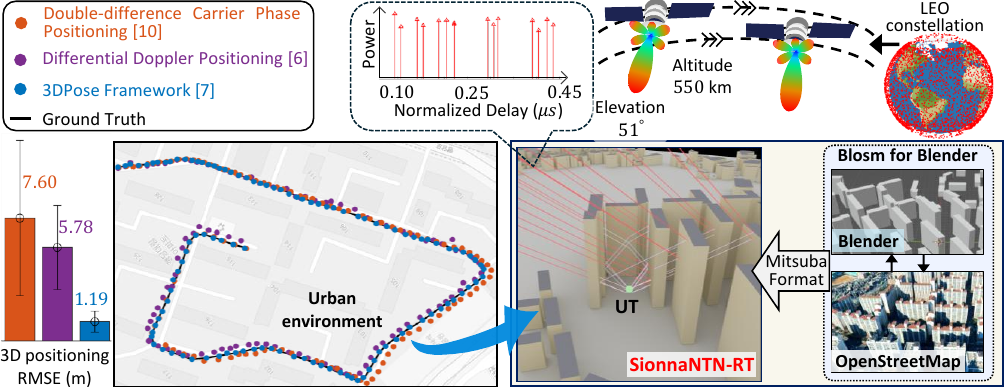}
     \caption{Evaluation map for urban environment with moving UTs. SionnaNTN-RT incorporates OpenStreetMap-derived 3D building models, manually processed in Blender with the Blosm and Mitsuba add-ons, to enable realistic ray-tracing-based propagation modeling.}
    \label{fig: map} 
\end{figure*}

\begin{table*}[t]\renewcommand{\arraystretch}{1.2}
    \centering
    \caption{Positioning performance comparison of Doppler and non-Doppler methods in different environments}
    \resizebox{\linewidth}{!}{
    \begin{tabular}{>{\hspace{-0.5em}}l  >{\hspace{-1.5em}}l >{\hspace{-2.5em}}c<{\hspace{-1em}} >{\hspace{-1em}}c >{\hspace{-0.8em}}c >{\hspace{-0.5em}}c >{\hspace{-0.8em}}c<{\hspace{-0.8em}}}
    \hlineB{3}
    & & & \multicolumn{4}{c}{3D positioning RMSE $\pm$ Standard deviation (m)} \\ \cline{4-7}
        & Positioning Methods &  & \multicolumn{2}{c}{Open-sky environment}  & \multicolumn{2}{c}{Urban environment}  \\ \cline{4-7}
           & & \multirow{-3}{*}{\begin{tabular}{>{\hspace{-0.8em}}c}
              Use of \\ Differential \\ Positioning 
         \end{tabular}}
         &  Static UT & Moving UT & Static UT & Moving UT \\ \hline 
         &GNSS pseudorange positioning \cite{morton2021position}
         & No &  $\num{2.424}~\pm~\num{1.057}$  & $\num{2.292}~\pm~\num{1.05}$  &  $\num{18.743}~\pm~\num{3.06}$  & $\num{28.254}~\pm~\num{19.22}$ \\ 
         \multirow{-2}{*}{\begin{tabular}{>{\hspace{-0.8em}}c}
              Non-Doppler \\ method 
         \end{tabular}}
         & \cellcolor[HTML]{EEF3FF} \begin{tabular}{>{\hspace{-1em}}c}\cellcolor[HTML]{EEF3FF}Double-difference carrier phase positioning \cite{9110199}\end{tabular}
          &\cellcolor[HTML]{EEF3FF}Yes &\cellcolor[HTML]{EEF3FF} $\num{3.520}~\pm~\num{1.83}$ &\cellcolor[HTML]{EEF3FF} $\num{2.443}~\pm~\num{1.19}$ &\cellcolor[HTML]{EEF3FF} $\num{4.850}~\pm~\num{2.39}$  &\cellcolor[HTML]{EEF3FF} $\num{7.606}~\pm~\num{4.81}$ \\ \hline 
         &  Non-differential Doppler positioning \cite{shi2023revisiting}
          & No & $\num{2228}~\pm~\num{76}$  & $\num{2228}~\pm~\num{77}$  & $\num{2305}~\pm~\num{134}$  & $\num{2305}~\pm~\num{135}$  \\ 
          \rowcolor[HTML]{EEF3FF}
         \cellcolor[HTML]{FFFFFF}
          & \cellcolor[HTML]{EEF3FF} \begin{tabular}{>{\hspace{-1em}}c}\cellcolor[HTML]{EEF3FF}Doppler positioning based on EKF \cite{BENZERROUK2019496}\end{tabular}
          & No & $\num{2160}~\pm~\num{79}$  & $\num{2163}~\pm~\num{79}$  & $\num{2281}~\pm~\num{171}$  & $\num{2281}~\pm~\num{172}$  \\  
         & Differential Doppler positioning \cite{dif9843493}
         & Yes & $\num{2.863}~\pm~\num{1.48}$  & $\num{3.925}~\pm~\num{1.09}$ & $\num{5.587}~\pm~\num{2.53}$  & $\num{5.786}~\pm~\num{2.61}$  \\ 
        \rowcolor[HTML]{EEF3FF}
         \cellcolor[HTML]{FFFFFF}
         \multirow{-4}{*}{\begin{tabular}{>{\hspace{-0.5em}}c}
             \cellcolor[HTML]{FFFFFF} Doppler \\ 
             \cellcolor[HTML]{FFFFFF} method 
         \end{tabular}}
        &\cellcolor[HTML]{EEF3FF} \begin{tabular}{>{\hspace{-1em}}c}\cellcolor[HTML]{EEF3FF}3DPose framework \cite{10716274}\end{tabular} & Yes & $\textbf{0.781}~\pm~\textbf{0.38}$  & $\textbf{0.803}~\pm~\textbf{0.40}$ & $\textbf{0.906}~\pm~\textbf{0.44}$  & $\textbf{1.195}~\pm~\textbf{0.65}$  \\
         \hlineB{3}
    \end{tabular}
    }
    \label{tab: com}
\end{table*}

{A positioning GUI simulator, shown in Fig.~\ref{fig: simulator}, provides an effective tool for evaluating the impact of ephemeris error on Doppler positioning.}
By simulating predefined ephemeris errors, the simulator generates measurement datasets for both Doppler shift and GNSS pseudorange positioning, displaying positioning outcomes in terms of RMSE for north, east, and down coordinates, along with time series error graphs.
Simulation results confirm that the 3DPose framework \cite{10716274} outperforms other methods in terms of accuracy and robustness. 
Fig.~\ref{fig: map} represents an evaluation map comparing the positioning performance of Double-difference carrier phase positioning \cite{9110199}, Differential Doppler positioning \cite{dif9843493}, and 3DPose framework \cite{10716274} in urban environments. The results indicate that Double-difference carrier phase positioning \cite{9110199} exhibits significant degradation due to multipath interference and NLOS effects in urban environments.

Table~\ref{tab: com} presents the 3D positioning RMSE for different methods across open-sky and urban environments considering both static and moving UTs.
Non-differential Doppler positioning \cite{shi2023revisiting} and Doppler positioning based on EKF \cite{BENZERROUK2019496} exhibit kilometer-level errors across all scenarios primarily due to ephemeris inaccuracies.
Doppler-based positioning is highly sensitive to ephemeris errors, which remain largely unmitigated by non-differential processing, while EKF primarily suppresses measurement noise.
In contrast, GNSS pseudorange positioning \cite{morton2021position} achieves significantly better accuracy, with RMSE values of $\num{2.424}$ m for static UTs and $\num{2.292}$ m for moving UTs in the open-sky. However, in an urban environment, where multipath effects are more pronounced, the RMSE of GNSS pseudorange positioning \cite{morton2021position} increases dramatically to $\num{18.743}$ m for static UTs and $\num{28.254}$ m for moving UTs, highlighting the susceptibility of GNSS pseudorange positioning to signal obstruction and reflection.

{To mitigate the impact of ephemeris errors on UT positioning, a base receiver with a known position is introduced for position correction.}
Double-difference carrier phase positioning \cite{9110199} achieves RMSE values of $\num{3.520}$ m for static UTs and $\num{2.443}$ m for moving UTs in the open-sky environment, increasing to $\num{4.850}$ m and $\num{7.606}$ m in the urban environment due to multipath effects.
Differential Doppler positioning \cite{dif9843493} exhibits RMSE values of $\num{2.863}$ m for static and $\num{3.925}$ m for moving UTs in open-sky conditions, while rising to $\num{5.587}$ m and $\num{5.786}$ m in urban environments.
However, by incorporating an ephemeris error correction algorithm, the 3DPose framework \cite{10716274} achieves significantly improved positioning accuracy, reducing the RMSE to $\num{0.781}$ m and $\num{0.803}$ m for static and moving UTs, respectively, under open-sky conditions. In the urban environment, it maintains high accuracy, achieving RMSE values of $\num{0.906}$ m for static UTs and $\num{1.195}$ m for moving UTs.

Overall, the findings highlight the potential of Doppler shift-based positioning with LEO satellites when supplemented with base receiver measurements over non-Doppler methods.
Implementing an ephemeris error correction algorithm further enhances the accuracy and reliability of UT positioning, demonstrating its feasibility as a viable alternative to conventional GNSS positioning methods.

\section{Future Research Directions}

Leveraging Doppler shift measurements from LEO satellites offers new possibilities for precise localization, particularly in challenging scenarios such as urban canyons and indoor settings. This section outlines key future research directions to enhance the effectiveness of LEO satellite-based positioning. 


\subsubsection{Satellite selection algorithm}

Due to the large scale of LEO mega-constellations, a considerable number of satellites may be simultaneously visible to the UT, increasing computational complexity while contributing unequally to positioning accuracy.
Geometric dilution of precision (GDOP) depends on the satellite-to-UT LOS unit vector, while DDOP accounts for both the LOS unit vector and satellite velocity direction. A low GDOP value does not always imply a good DDOP value. 
Therefore, a dedicated selection algorithm based on DDOP should be designed to identify satellites that maximize positioning accuracy, minimize computational overhead, and ensure the scalability of LEO-based positioning.

\subsubsection{Doppler shift tracking from multiple satellites}

Tracking Doppler shifts from multiple satellites presents another challenge for practical deployment and scalability.
The proprietary signal structures of the LEO satellites are unlikely to be disclosed for public use, making signal tracking particularly challenging.
In addition, different LEO satellites may employ distinct transmission frequencies and unique signal structures. 
A robust and efficient Doppler tracking algorithm should be developed to estimate Doppler shifts from more than seven satellites simultaneously, ensuring reliable positioning.

\subsubsection{Error source modeling for LEO satellites}

Doppler-based precise positioning requires accurate satellite time and orbital information, atmospheric delay rate corrections, and mitigation of other measurement errors.
Base receivers with known positions can partially correct orbital errors, but there is currently no dedicated atmospheric model for LEO satellites. 
Specifically, the existing ionospheric model is primarily designed for GNSS, necessitating further research into LEO-specific atmospheric correction models.
{Additionally, while Doppler shift measurements are less susceptible to multipath error than pseudorange measurements, their performance in a dense urban environment requires further investigation.}
The proposed \mbox{SionnaNTN-RT} ray-tracing framework could serve as a feasible tool for this purpose.

\subsubsection{Integrating Doppler shift measurements with 3GPP NTN}

The current 3rd Generation Partnership Project (3GPP) standard verifies the user equipment location with a dedicated position reference signal (PRS) using the multi-round trip time (Multi-RTT) \cite{10355106}. 
Incorporating Doppler shift-based positioning into the 3GPP standard could significantly improve localization performance, particularly in LOS-blocked regions.
{In addition, LEO Doppler positioning enables NTN D2D to be GNSS-free.}
By leveraging dual communication capabilities, UTs could achieve precise indoor positioning by utilizing signals from LEO satellites, thereby expanding the applicability of satellite-based navigation in urban and indoor settings. 


\section{Conclusion and Outlook}
\label{sec: Conclusion}
Doppler shift-based positioning is emerging as a viable approach for SOPs-based localization using large-scale LEO satellite constellations.
This article has presented the fundamental principles, key challenges, and potential advantages of exploiting Doppler information over conventional methods. Simulation results demonstrated the effects of ephemeris and multipath errors, highlighting the potential of Doppler-based techniques to achieve reliable positioning in both open-sky and urban environments. 
Furthermore, the proposed \mbox{SionnaNTN-RT} ray-tracing framework is a deterministic tool for modeling realistic signal propagation environments.
Finally, we have outlined potential future research directions to advance Doppler shift-based positioning. 
As LEO satellite networks continue to expand, Doppler-based positioning is expected to play a pivotal role in delivering accurate and resilient localization solutions across a wide range of applications.

\bibliographystyle{IEEEtran}
\bibliography{ref}

\end{document}